\documentclass[12pt,titlepage]{article}
\usepackage{epsfig}
\def\L{{\bf L}}
\def\F{{\bf F}}
\def\GU{{\Gamma_0(2)}}
\def\G2{{\Gamma(2)}}
\def\Ph{{\bf P}}
\def\PH{{\bf P_2}}
\begin{document}
\title{Duality in the Quantum Hall Effect - the R\^ole of Electron Spin}
\author{{Brian P. Dolan}\\
{\small Department of Mathematical Physics, National University
of Ireland, Maynooth, Ireland}\\
{\small and}\\
{\small Dublin Institute for Advanced Studies,
10 Burlington Rd., Dublin, Ireland}\\
{\small e-mail: bdolan@stp.dias.ie}}
\date{15th February 2000}
\maketitle
\begin{abstract}
At low temperatures the phase diagram for the quantum Hall effect has a powerful symmetry
arising from the Law of Corresponding States.
This symmetry gives rise to
an infinite order discrete group which is a generalisation of Kramers-Wannier
duality for the two dimensional Ising model. The duality group, which is a subgroup
of the modular group, is
analysed and it is argued that there is a quantitative difference 
between a situation in which
the spin splitting of electron energy levels is comparable
to the cyclotron energy and one in which the spin splitting is much less than the
cyclotron energy. In the former case the group of symmetries is larger than in the latter 
case. These duality symmetries are used to constrain the scaling functions of the
theory and, under an assumption of complex meromorphicity, a unique functional
form is obtained for the crossover of the conductivities
between Hall states as a function of the external magnetic
field. This analytic form is shown to give good agreement with experimental data.

The analysis requires a consideration of the way in which longitudinal resistivities
are extracted from the experimentally measured longitudinal resistances and a novel
method is proposed for determining the correct normalisation for the former.

\bigskip
\noindent\hbox{Report No. DIAS-STP-00-01}\hfill \hbox{PACS nos: 73.40.Hm, 05.30.Fk, 02.20.-a}
\end{abstract}
\pagebreak
\bigskip
\section{Introduction}

The current understanding of the details of quantum Hall effect relies
crucially on the composite fermion picture \cite{Jain}.
An important step in the development of this understanding was the introduction
of statistical gauge fields and the r\^ole they play
in the Landau-Ginsparg effective action, \cite{ZHK}. In fact the Law of Corresponding States
of reference \cite{KLZ} has an interpretation as a beautiful and extremely
powerful generalisation of Kramers-Wannier duality, \cite{Lutken+RossA} \cite{Lutken+RossB},
though references \cite{KLZ} and \cite{Lutken+RossA} were completely independent
of one another and appeared almost simultaneously.

 Kramers-Wannier duality is a discrete ${\bf Z}_2$ map of the partition function
for the 2-dimensional Ising model at one temperature to the same partition function at
a different temperature. Its power lies in the prediction of the critical temperature
without having to solve the model. Since its discovery however other techniques
have proven more useful --- not only because the model has since been solved exactly
in 2-dimensions but also because, being a discrete symmetry, duality is all or nothing affair.
There is no algorithm for finding such symmetries --- it is hit and miss guesswork.
This is one of the reasons why the renormalisation group approach has proven more 
powerful as a technique for understanding the physics at and near critical points
associated with second order phase transitions ---
not only for the 2 and 3-dimensional Ising models in the presence of an external magnetic
field but also for many other models. The renormalisation group (RG) approach cannot
really be called an algorithm (it has many different manifestations all of which
come under the same general heading of ``RG'') but it is nevertheless algorithmic
in nature --- given a field theoretical model one can perturb around free field
theory and hope to apply some version of the RG to extract physical information
without actually having to solve the model.

 Nevertheless interest in duality has never died perhaps because of the elegance of the
idea but also because, if one {\it can} find a duality, one can obtain exact results
whereas the perturbative RG is always an approximation. With duality the chances of
success are smaller but the payoff is greater. 
A ${\bf Z}_2$ duality for massive Maxwell-Chern-Simons theory was discovered
in \cite{Rey+Zee} which necessitated the introduction of a complex coupling
with positive imaginary part.
A generalisation of Kramers-Wannier
duality from ${\bf Z}_2$ to an infinite discrete non-Abelian group (called
the modular group, which is defined in the next section) 
was discovered for a coupled clock model in \cite{Cardy}. Being a larger group
this leads to a much more complicated phase diagram and it has been suggested that
a version of this duality applies to Potts models with a complex temperature,
\cite{Bertoldi}. 
The importance of the modular group, and some of its sub-groups, for
the quantum Hall effect was realised in \cite{Lutken+RossA} and \cite{Lutken+RossB}
and this led to the development of the phase diagram compatible
with modular symmetry first presented in \cite{Lutkenb}.
Interest in duality has also seen a major revival recently in the
context of super-symmetric gauge theories and string theory (for a review see 
\cite{DualityReview}). 

 Very strong results can be obtained by combining duality symmetry (when one can
be derived) with RG techniques. This was first done for the quantum Hall effect in
\cite{Burgess+Lutkena}. The basic idea is that the duality map should commute with the RG
flow which puts very strong constraints on the $\beta$-functions of the theory.
For example fixed (i.e. self-dual) points of duality symmetry
correspond to fixed points of the RG and therefore the $\beta$-functions would be
expected to vanish at these points --- even if they correspond to strongly
coupled theories in the field theoretic sense. Since \cite{Burgess+Lutkena} appeared
a number of authors has examined the constraints put on the RG flow by modular
symmetry under various other assumptions, 
\cite{BDa},\cite{BDb},\cite{Taniguchi},\cite{Burgess+Lutkenb},\cite{GWT-Mb}.
Indeed one can use this approach, together with particle-hole symmetry, 
to derive the semi-circle law for transitions for
quantum Hall transitions very generally without any recourse to a specific
microscopic model --- it follows very generally from the duality symmetry, \cite{BDD}.

 The aim of this paper is to explore further the consequences of this infinite
generalisation of Kramers-Wannier duality for the quantum Hall effect. In particular
electron spin is incorporated into the duality picture and it
is argued that when the splitting of the Landau levels due to
electron spin is small compared to the cyclotron
energy the relevant duality symmetry is smaller than when the
Landau levels are all well separated. Technically speaking the symmetry group is $\G2$ in the
former case and $\GU$ in the latter (these groups are defined in the next section).

 In order to compare this prediction with experiment some extra assumptions about
the form of the $\beta$-functions are necessary and here the analysis of \cite{BDb}
is applied to the case of small spin splitting. This involves a very
specific assumption concerning the form of the $\beta$-functions --- that they are
meromorphic in the sense of a natural complex conductivity, $\sigma:=\sigma_{xy}+i\sigma_{xx}$
(throughout this paper units are used in which $e^2/h=1$). Unfortunately this assumption has
no microscopic justification to date, rather it is made on the basis of analogy with
other models with modular symmetry as a duality, \cite{SW}. However it leads
to such specific predictions that is easily falsifiable, though comparisons with experimental
data show that its predictions agree very well with the currently available data lending
some confidence that a justification may be found in the future. The 
available data to date, at strong fields and at low enough temperatures for the
Law of Corresponding States to represent a symmetry, seem to be compatible with $\GU$
when the meromorphic ansatz is used,
so it is not possible to test the specific predictions
made here concerning the $\G2$ case.  This must wait
for future experiments.

 An important aspect of the comparison with experimental data made
in section six is the normalisation of the longitudinal resistivity, $\rho_{xx}$.
This is a problem for experiment, which in reality measures the longitudinal
resistance $R_{xx}$, and it is usually assumed these are related by
$R_{xx}={L\over W}\rho_{xx}$ in a rectangular
sample of length $L$ and width $W$. The assumptions that go into this relation are
examined in section five where it is argued that a better formula is
$R_{xx}={L\over W}f(W/L)\rho_{xx}$ where $f(W/L)$ is an undetermined 
sample dependent function
with the universal property that $\lim_{W/L\rightarrow 0}f(W/L)=1$.
The duality symmetry of the quantum Hall effect however suggests an alternative
method for extracting $\rho_{xx}$ from the experimental $R_{xx}$, at least
for Hall-Hall transitions. Given that
the duality group, together with particle-hole symmetry, predicts the semi-circle law
one can use this to determine the normalisation of $R_{xx}$ by {\it assuming}
the semi-circle law at sufficiently low temperatures. This technique cannot determine
the normalisation by using a Laughlin state-insulator transition because these correspond
to vertical lines in the complex $\rho$-plane, which remain vertical under any re-scaling
of $\rho_{xx}$. However this very fact dictates that Laughlin state-insulator crossovers
should be semi-circles in the
$\sigma$-plane for any normalisation of $\rho_{xx}$ 
and this gives a quantitative method of determining when the temperature
is `low enough' for the infinite duality symmetry to be valid --- the Laughlin 
state-insulator transitions must be semi-circles in the complex $\sigma$-plane.
When this is the case the normalisation of $\rho_{xx}$ for a Hall-Hall transition
can be extracted from the experimental data by choosing the normalisation so as to
get as close as possible to a semi-circle for the crossover in the complex $\rho$-plane.

 The layout of the paper is as follows: in section two the definition of the
modular group and its various relevant sub-groups is reviewed together with
its action in the quantum Hall effect; section three discusses the relation between
the modular group and scaling; section four introduces the specific ansatz
of meromorphic $\beta$-functions; section five is devoted to the relation between
the longitudinal resistance and resistivity while section six compares the predictions
of section four to existing experimental data. Finally section seven contains
a summary and conclusions.

\section{The Modular Group}
The law of corresponding states \cite{KLZ} was originally proposed on the basis
of an effective field theory --- Maxwell-Chern-Simons theory. 
It provides a powerful method for classifying
quantum Hall states and transitions
between them, both integral and fractional, as well as for Hall-insulator
transitions.
The law is expressed by writing the conductivities (both longitudinal $\sigma_{xx}$
and transverse $\sigma_{xy}$) as functions of the filling factor $\nu$. Then
the following transformations were defined in \cite{KLZ}:

\begin{itemize}
\item
Landau Level Addition Transformation ({\bf L})
$$
\sigma_{xy}(\nu +1)\leftrightarrow \sigma_{xy}(\nu) + 1,\quad
\sigma_{xx}(\nu +1)\leftrightarrow \sigma_{xx}(\nu)
$$
\item
Flux Attachment Transformation ({\bf F})
$$
\rho_{xy}\left({\nu\over 2\nu +1}\right)
\leftrightarrow \rho_{xy}(\nu) +2,\quad
\rho_{xx}\left({\nu\over 2\nu +1}\right)
\leftrightarrow \rho_{xx}(\nu)
$$
\item
Particle-Hole Transformation ({\bf P})
$$
\sigma_{xy}(1-\nu)\leftrightarrow 1-\sigma_{xy}(\nu),\quad
\sigma_{xx}(1-\nu)\leftrightarrow \sigma_{xx}(\nu).
$$
\end{itemize}
As discussed in \cite{KLZ} under certain circumstances 
the arrows above can be replaced by equalities
and then these transformations
become {\it symmetries}. This is not expected to be true in general
but should hold, for example, as the temperature $T$ tends to zero.

At almost exactly the same time, but from a completely different perspective,
the modular group \cite{Lutken+RossA} or some of its sub-groups \cite{Lutken+RossB}
were also implicated as being relevant to these phenomena. 
In this approach one defines a {\it complex} conductivity, $\sigma:=\sigma_{xy}
+i\sigma_{xx}$, which by necessity lives on the upper-half complex plane since
$\sigma_{xx}\ge 0$, and a group action
\begin{equation}
\label{gammadef}
\gamma(\sigma):={a\sigma +b \over c\sigma +d},
\end{equation}
where $a,b,c$ and $d$ are any four integers satisfying $ad-bc =1$.
This condition can re-written by defining a $2\times 2$ matrix
$\gamma=
\left(
\matrix{a&b\cr c&d\cr}
\right)$
and demanding that $\det\gamma=1$. It is easy to check from the definition
(\ref{gammadef}) that, for any three
such matrices satisfying $\gamma_1\gamma_2=\gamma_3$, 
we have $\gamma_1(\gamma_2(\sigma))=\gamma_3(\sigma)$. Thus the group multiplication
law is given by matrix multiplication.
Note that for any given matrix $\gamma$ the matrix $-\gamma$ gives the same transformation
since the minus signs cancel above and below in (\ref{gammadef}).
The resulting group is called the modular group in the mathematical literature,
sometimes denoted $\Gamma(1)$, and it is related to 
$Sl(2,{\bf Z})$.\footnote{That is the sub-group of
the group of $2\times 2$ matrices of real numbers
with unit determinant ($Sl(2,{\bf R})$) obtained by restricting to integral entries.}
They are not quite the same group because $Sl(2,{\bf Z})$ distinguishes between
$\gamma$ and $-\gamma$ whereas $\Gamma(1)$ does not, so $\Gamma(1)$ is obtained from
$Sl(2,{\bf Z})$ by identifying $\gamma$ and $-\gamma$ as group elements, or in other
words projecting out by a ${\bf Z}_2$ factor, so 
$\Gamma(1)=PSl(2, {\bf Z}):=Sl(2, {\bf Z})/{\bf Z}_2$.
Any element of the modular group
can be obtained by taking a string of products of the two generators
$\left(
\matrix{1&1\cr 0&1\cr}
\right)$ and 
$
\left(
\matrix{0&1\cr\ -1&0\cr}
\right)$.

In this matrix notation Landau level addition and Flux attachment are represented
by
$\L=
\left(
\matrix{1&1\cr 0&1\cr}
\right)$ and $\F=
\left(
\matrix{1&0\cr 2&1\cr}
\right)$
and these two transformations generate an infinite discrete group,
which is a sub-group of $\Gamma(1)$, often
denoted by $\Gamma_0(2)$ in the mathematical literature (see \cite{Koblitz}
for example, another notation commonly used is $\Gamma_U(2)$, \cite{Rankin}).
Any element $\gamma\in {\GU}$ can be represented by some string of products of
$\L$ and $\F$.
This sub-group is most succinctly described by restricting the integer $c$ in
the matrix $\gamma$ to be even.
The particle-hole transformation ${\Ph}(\sigma)= 1-\bar\sigma$ is
an outer auto-morphism of the modular group and also of the sub-group, $\GU$.

The hypothesis that these
transformations represent a symmetry can now be expressed as
\begin{eqnarray}
\sigma(\nu+1)=&{\bf L}(\sigma(\nu))= \sigma(\nu) + 1\qquad
\sigma\Bigl({\nu\over 2\nu +1}\Bigr)={\bf F}(\sigma(\nu))={\sigma(\nu)\over 2\sigma(\nu)+1}\\
&\sigma(1-\nu)={\Ph}(\sigma(\nu))= 1-\bar\sigma(\nu)
\end{eqnarray}
(flux attachment is easily transcribed from resistivities to conductivities,
in complex notation,
by $\sigma=-1/\rho$, with $\rho=-\rho_{xy}+i\rho_{xx}$).

Other sub-groups of $\Gamma(1)$ have been proposed in the literature as
being relevant to the quantum Hall effect, \cite{Lutkena}, \cite{GWT-Ma}, \cite{Wilczek+Shapere}
and it shall be argued here  that the group proposed in \cite{GWT-Ma} is relevant when spin
splitting of the electrons results in the Landau levels that are not very
well separated. Obviously the Zeeman effect would result in Landau levels that are
split but it is also argued in \cite{Fogler} the exchange energy 
between electrons of opposite spin in electron-electron
interactions can be larger than the Zeeman energy.

Thus, despite the
difference in philosophy, the two approaches in \cite{KLZ} and \cite{Lutken+RossB}
give the same results
at zero temperature, where the law
of corresponding states becomes a symmetry in \cite{KLZ} which is precisely
the symmetry group $\Gamma_0(2)$ in \cite{Lutken+RossB}. The latter approach is
in many ways more powerful however since, once one realizes that the law of corresponding
states is a group, one can bring the full machinery of group theory to bear
on the problem --- indeed there is a very highly developed and beautiful mathematical
literature on the modular group and its sub-groups and the transformation properties
of functions of $\sigma$ under this group action, \cite{Koblitz} \cite{Rankin}.
Thus, for example, one can deduce the selection rule $\vert p_1q_2-p_2q_1\vert=1$
for transitions between two states with filling factors $\nu_1=p_1/q_1$
and $\nu_2=p_2/q_2$, with $q_1$ and $q_2$ odd, \cite{BDa}.

However the law of corresponding states does serve to highlight the
physics of the problem --- in particular the Landau level addition symmetry
assumes that the physics at any partially filled Landau level is independent of how
many lower Landau levels are completely full. While this seems a reasonable
assumption when the Landau levels are well spaced it is not so reasonable
if any two Landau levels get close to one another. 
Spin splitting is important here:
if one ignores electron spin the Landau levels of the unperturbed Hamiltonian
are all evenly spaced, but including spin and a Land\'e $g$-factor could result
either in well spaced Landau levels (if the Landau splitting is comparable to
the cyclotron energy) or 
in Landau levels occurring in pairs which are close to one another
(if the spin splitting is very small). 
Some possibilities are shown in figure 1.
The important parameter here is the ratio of the spin splitting energy to the
cyclotron energy. If this ratio is precisely $1/4$ then all the split Landau levels
are evenly spaced --- at exactly half the cyclotron energy of the spinless case.
Experimentally the value is often near $1/4$, \cite{Fogler}.
However if the ratio of the spin splitting energy to the cyclotron energy
is either very small or near a half (more generally near an integer or a half integer),
then the Landau levels lie
very close to each other in pairs.
In this latter case the physics in the upper member of a pair 
could well be influenced by the presence
of electrons in its lower member. However, when the pairs are well separated
it would seem reasonable to assume that the physics of any {\it pair} is independent
of how many pairs below are filled.
This is tantamount to replacing the Landau level addition transformation
with $\L^2$:
\begin{equation}
\sigma(\nu+2)={\L^2}(\sigma(\nu))= \sigma(\nu) + 2.
\end{equation}

There does not seem to be any reason to expect flux attachment to be modified
when the spin splitting is small
and we shall therefore concentrate on the group generated 
by $\F$ and $\L^2$. This sub-group of $\GU$ is usually denoted by $\G2$ in the
mathematical literature and it can can also be described by demanding that
both $b$ and $c$ in (\ref{gammadef}) be even integers.
This group has already received some attention in relation to the quantum Hall
effect, \cite{Lutkena},\cite{GWT-Ma}.

It is our intention to explore the differences between $\GU$ and $\G2$, 
as applied to the quantum Hall effect and to the scaling flow of the
conductivities in particular. To this end the next section deals with
the relation between the modular group and scaling.
 
\section{The Modular Group and Scaling}
Very powerful predictions can be made by combining the two parameter scaling hypotheses
of Khmel'nitskii \cite{Khmel}, as implemented in \cite{Pruisken},
with modular symmetry of the quantum Hall effect. One assumes that there is
a physical length scale, $L$, upon which the conductivities depend --- $L$ could
be the electron scattering length, for example, which diverges as $T\rightarrow 0$
in an infinite system.  It is argued in \cite{Pruisken} that $L$
should depend on the external magnetic field $B$ and the temperature $T$ through
a single scaling variable $L(\Delta B/T^\mu)$, where $\Delta B=B-B_c$ is the
deviation of the external magnetic field from the critical value $B_c$ 
and $\mu$ is an anomalous dimension.
The derivatives
\begin{equation}
\label{betaL}
\beta^L_{xy}:=L{d\sigma_{xy}\over dL}\qquad\beta^L_{xx}:=L{d\sigma_{xx}\over dL}
\end{equation}
give flow equations for the conductivities, whose integral curves,
$\sigma_{xy}(\Delta B/T^\mu)$ and $\sigma_{xx}(\Delta B/T^\mu)$, 
are the scaling functions of \cite{Pruisken}. 
More generally one would expect the conductivities to depend on more parameters,
such as the density of impurities (integer effect)
or the charge carrier density (fractional effect).  Scaling arguments
would then lead to the form $\sigma_{xy}(\Delta B/T^\mu,n/T^\zeta)$ and
$\sigma_{xy}(\Delta B/T^\mu,n/T^\zeta)$ for some exponent $\zeta$,
where $n$ is the charge carrier density, for example. 
Thus varying $B$ and $T$ independently,
while keeping $n$ fixed, gives two free parameters with which to explore the
whole upper-half $\sigma$-plane. Alternatively one could keep $B$ fixed
and vary $T$ and $n$. Pruisken's scaling form is recovered at low temperatures,
provided the conductivities are well behaved functions of the second argument,
which tends to zero as $T\rightarrow 0$ if
the exponent $\zeta$ is negative and diverges as $T\rightarrow 0$ if
$\zeta$ is positive. In the following it shall be assumed that $T$ is low enough
that the second argument here can be ignored and the conductivities are functions
of a single scaling variable, $b:={\Delta B\over T^\mu}$.

More generally one could define $\beta$-functions to be
derivatives with respect to a function of $b$,
or alternatively one
could use filling factors, $\nu={ne\over B}$, and consider 
derivatives with respect to a function $s(v)$ of
$v:={ne\over T^\mu}\Delta\nu$, with $\Delta\nu:=\nu - \nu_c$,
\begin{equation}
\label{betas}
\beta_{xy}:={d\sigma_{xy}\over ds}\qquad\beta_{xx}:={d\sigma_{xx}\over ds}.
\end{equation}
These latter are clearly not the same as the $\beta$-functions in (\ref{betaL}).
Indeed it may be an abuse of language to call them $\beta$-functions --- the
integrated form $\sigma_{xx}(v)$ and $\sigma_{xy}(v)$ might be more correctly
called scaling functions\footnote{I am grateful to Shivaji Sondhi for discussions
on this point.}--- but we shall continue to use the notation of equation~(\ref{betas})
since these functions are somewhat analogous to Callan-Symanzik functions of
relativistic quantum  field theory.
One of the principle assumptions used in the following is that the
action of $\GU$ (or $\G2$) should commute with the flow equations~(\ref{betas}),
\cite{Burgess+Lutkena}.

When spin splitting leads to well separated Landau levels
($\GU$ case) one can make predictions
about the global phase diagram in the complex $\sigma$-plane and the positions
of the critical points in the crossover between two Hall plateaux and between
a Laughlin state and the insulator,
\cite{Lutkenb}, \cite{BDa}, \cite{Lutkena}.
Moreover one can deduce the
semi-circle law, \cite{BDD}, without making any further assumptions ---
indeed without recourse to any specific microscopic model. 
This law was proposed on the basis of an analysis of a specific microscopic model
in \cite{Ruzin} and has strong experimental support \cite{Hilkeetal}.
There is also a duality for any transition
between two Hall plateaux 
at
$\nu_1=p_1/q_1$
and $\nu_2=p_2/q_2$, with $q_1$ and $q_2$ odd.\footnote{This includes the case of a Laughlin state--insulator transition with $\nu_1=1/q$ and 
$\nu_2=0$ ($p_1=q_2=1$, $q_1=q$ and $p_2=0$).
In the sequel any mention of a Hall-Hall transition is implied to include the
Laughlin state-insulator transition, unless otherwise stated explicitly.} 
This duality is
given by parameterizing the
semi-circle by $0\le w<\infty$ in
$\sigma={(q_1p_1+w^2q_2p_2)+iw\over (q_1^2 + w^2 q_2^2)}$ (where $p_1q_2-p_2q_1=1$)
and inverting $w\rightarrow 1/w$.
For the Laughlin state--insulator transition, this $w\rightarrow 1/w$
inversion leads to the observed experimental duality $\rho_{xx}\rightarrow 1/\rho_{xx}$ under
$\Delta B\rightarrow -\Delta B$
\cite{STSCSS}.
In particular a general formula for the values of the conductivities
at the critical points is given by setting $w=1$,
$\sigma_c={p_1q_1+p_2q_2 +i \over q_1^2 + q_2^2}$, \cite{BDb} --- these
are fixed points of $\GU$ in the sense that there exists an element
of $\GU$ that leaves these points invariant. They are the only such
points above the real axis in the complex $\sigma$-plane.

By making further assumptions about the form of the scaling functions one
can make much more specific predictions about the form of the crossover
between two Hall states for the $\GU$ case,
\cite{BDb},\cite{Taniguchi},\cite{Burgess+Lutkenb}.

For the $\G2$ case, when spin splitting does not separate the Landau levels completely,
there has been an analysis of the possible scaling flow too, \cite{GWT-Mb},\cite{Lutkena}.
Though in this case there is no prediction of the critical conductivities
for any transitions  because the group $\G2$ has no fixed
points above the real axis. However one can still derive the semi-circle law 
using exactly the same techniques as in \cite{BDD}, except applied to
$\G2$ rather than $\GU$, provided one assumes a generalisation for particle-hole
symmetry for $\G2$,
\begin{equation}
\label{PH2}
\sigma(2-\nu)={\PH}(\sigma(\nu)):=2-\bar\sigma(\nu).
\end{equation}

To go further in the analysis of a flow compatible with $\G2$ we need to make
further assumptions about the form of the $\beta$-functions in~(\ref{betas})
and in the next section the assumptions and techniques adopted in \cite{BDb}
for the $\GU$ case will be adapted and modified to further the analysis
of \cite{GWT-Mb}.

\section{Analytic $\beta$-functions}

In terms of the complex conductivity, equation~(\ref{betas}) can be
written as
\begin{equation}
\label{generalbeta}
{d\sigma\over ds}=\beta(\sigma,\bar\sigma)=\beta_{xy}+i\beta_{xx}.
\end{equation}
Demanding that the flow described in~(\ref{generalbeta}) commute with
the action of $\GU$ or $\G2$ requires that \cite{Burgess+Lutkena}
\begin{equation}
\label{modform}
{d\gamma(\sigma)\over ds}={1\over (c\sigma + d)^2}{d\sigma\over ds}.
\end{equation}
This equation follows easily from (\ref{gammadef}) and the fact that
$\det\gamma=1$.
To go any further requires making some further assumptions. Various
possibilities have been suggested in the literature,
\cite{BDb} --- \cite{GWT-Mb}, and the line we shall follow is to
investigate further the consequences of the assumption that the $\beta$-functions are
{\it meromorphic} functions of $\sigma$. Unfortunately this assumption has
no microscopic justification at the moment, rather it is made on the
basis of experience with super-symmetric QCD where the groups $\GU$ and
$\G2$ also appear and the $\beta$-function for $N=2$ super-Yang-Mills theory
are meromorphic functions of a complex coupling with
positive imaginary part, \cite{SW} \cite{Ritz}. There is certainly no evidence for
any real connection between the quantum Hall effect and super-symmetric
QCD, beyond the observation that the same discrete groups
crop up as symmetries of the physics in both cases, but it is interesting to apply
expertise gained from the occurrence of these groups in the latter framework
to the former, just to see what happens. The occurrence of meromorphic $\beta$-functions
in super-symmetric QCD is intimately related to super-symmetry and it is possible
that super-symmetry may have a r\^ole to play in the quantum Hall effect, \cite{Zirnbauer},
but this is as yet unclear. However it may be worth noting in passing that the modular
group was discovered in a statistical mechanical model prior to its use in the
quantum Hall effect \cite{Cardy} --- indeed this work analysed a model which was
motivated by ordinary QCD (i.e. not super-symmetric). 

A second motivation for the meromorphic assumption, 
and perhaps more important from a practical point of view, is that there is a very
highly developed mathematical literature on the theory of meromorphic
functions satisfying equation~(\ref{modform}) --- such functions are called
modular forms of weight $-2$ in the mathematical literature --- and this puts
very strong restrictions on the form of the $\beta$-functions. Thus even though
the meromorphic assumption lacks any physical foundation the predictions are extremely strong
and thus the assumption is easily falsifiable by experiment --- an important
aspect of any theoretical analysis. 
We shall see, in fact, that experiment agrees surprisingly
well with the consequences of this assumption.

For the $\GU$ case it is shown in \cite{BDb} that $\beta$-functions
satisfying 
\begin{equation}
\beta(\gamma(\sigma))={1\over (c\sigma +d)^2}\beta(\sigma),
\end{equation}
(with $c$ even) and subject to some further reasonable physical restrictions,
must have a precise analytic form in terms of well known classical functions
called Jacobi $\vartheta$-functions, \cite{WW}.
The further restrictions are: that the $\beta$-functions should vanish at $\sigma=1$
and $\sigma=2$ as fast as possible in order to explain the fantastic stability
of the Hall plateaux, which must be attractive fixed points of the flow;
that the $\beta$-functions should be finite as $\sigma_{xx}\rightarrow\infty$ --- the
weak coupling regime of field theory models; and finally that there are no
fixed points other than those mentioned above and those representing
the critical point in the transition between any two plateau.
A further assumption used in \cite{BDb}, that $\sigma_{xy}$ should be constant
as $\sigma_{xx}\rightarrow\infty$, is in fact not necessary since this is actually
a consequence of the other assumptions, \cite{BDD}.

The Jacobi $\vartheta$-functions are defined either as infinite sums 
or as infinite products in the variable $q:=e^{i\pi\sigma}$, which converge
for $\Im\sigma>0$:
\begin{eqnarray}
\vartheta_2 =& 2\sum_{n=0}^\infty q^{(n+{1\over 2})^2}&=
2q^{1\over 4}\prod_{n=1}^\infty\bigl(1-q^{2n}\bigr)\bigl(1+q^{2n}\bigr)^2\\
\vartheta_3 =& \sum_{n=-\infty}^\infty q^{n^2}&=
\prod_{n=1}^\infty\bigl(1-q^{2n}\bigr)\bigl(1+q^{2n-1}\bigr)^2\\
\vartheta_4 =& \sum_{n=-\infty}^\infty (-1)^nq^{n^2}&=
\prod_{n=1}^\infty\bigl(1-q^{2n}\bigr)\bigl(1-q^{2n-1}\bigr)^2,
\end{eqnarray}
and satisfy the identity 
\begin{equation}
\label{thetaidentity}
\vartheta_3^4=\vartheta_2^4+\vartheta_4^4.
\end{equation}
There is a very special meromorphic function that is 
{\it invariant} under $\GU$ and is the simplest
such function, in the sense that it has the fewest possible number of zeros and poles:
$f(\sigma):=-{\vartheta_3^4\vartheta_4^4\over\vartheta_2^8}$ satisfies 
$f(\gamma(\sigma))=f(\sigma)$ for any $\gamma\in\GU$.

The unique form of the $\beta$-function (up to a positive constant) 
that is compatible with the above assumptions is obtained in 
\cite{BDb} and is
\begin{equation}
\label{analyticbeta}
\beta(\sigma)=-{f\over f^\prime}
\end{equation}
where $f^\prime={df\over d\sigma}$. Of course to make any contact with
physics we still have to specify
the unknown function $s(v)$ that is used to define the $\beta$-functions
in (\ref{betas}) --- we shall say more about this later.

These ideas were applied to $\beta$-functions compatible with $\G2$
in \cite{GWT-Mb} but the resulting flow had no evidence of any critical point
point in the crossover between two plateau. 
We shall therefore re-examine
this flow, with a view to introducing such a critical point. To do this recall
the physical picture that $\GU$ is the relevant symmetry group when the
Landau levels are all well separated. Now imagine varying the relative
magnitudes of the spin splitting and the cyclotron energy,
for example by varying the Land\'e $g$-factor or by varying the effective
mass of the charge carriers, until the Landau levels are paired off
into close pairs which are well separated from other pairs, as in figure 1.
In this latter case the relevant group is $\G2$.
We thus have a continuous deformation from $\GU$ symmetry to $\G2$
symmetry, and it should be possible to deform the $\beta$-functions
for $\GU$ continuously to those for $\G2$.

The analysis in \cite{GWT-Mb} 
requires using the function $\lambda(\sigma)={\vartheta_2^4\over \vartheta_3^4}$
rather than $f$, since $\lambda$ is the appropriate invariant
function for $\G2$ --- $f$ is certainly invariant under $\G2$, since
$\G2\in\GU$ and $f$ is invariant under $\GU$, but it is not the
simplest such function. Of course since $f$ is invariant under $\G2$ 
it can be expressed in terms
of $\lambda$, explicitly $f={\lambda-1\over\lambda^2}$ where (\ref{thetaidentity})
has been used --- but one
cannot express $\lambda$ in terms of $f$ without introducing branch cuts.
One can thus write the $\GU$ symmetric $\beta$-functions (\ref{analyticbeta})
in terms of $\lambda$,
\begin{equation}
\beta(\sigma)={\lambda(1-\lambda)\over \lambda^\prime(2-\lambda)}.
\end{equation}
This has a pole in $\lambda$ at $\lambda=2$, which corresponds to $\sigma=(1+i)/2$
for the transition between the integer quantum Hall state at $\sigma=1$ and the
insulator at $\sigma=0$. Of course this pole is reflected at every image of $(1+i)/2$
under the action of $\GU$ and thus all the critical points for every quantum Hall transition
are generated from this one in the transition $\sigma=1\rightarrow\sigma=0$ by the
symmetry group. 
As discussed at length in \cite{BDb} this pole is not
a disaster, but is a perfectly regular manifestation of the critical point
in the crossover, at $\sigma_{xx}=1/2$ --- this will be discussed in more detail
a little later. The pole at $\sigma=(1+i)/2$ is a feature specific to $\GU$, 
it must be there because $(1+i)/2$ is a fixed point of $\GU$, but there is nothing about
$\G2$ that dictates the position of the critical point. It would therefore seem 
natural to postulate the $\G2$ $\beta$-functions to be the following deformation of the $\GU$
$\beta$-functions:
\begin{equation}
\label{betaG2}
\beta(\sigma)={\lambda(1-\lambda)\over \lambda^\prime(a-\lambda)},
\end{equation}
where $a$ is a sample specific constant. For well separated, spin split Landau
levels $a=2$ giving the $\GU$ $\beta$-functions (\ref{analyticbeta}), 
but when pairs of Landau levels lies close to each other $a$ can deviate from the
more symmetric value of $2$. If one assumes particle-hole symmetry $\PH$ of 
equation (\ref{PH2}) the semi-circle law follows and
the critical point must lie on the semi-circle spanning $0$ and $1$.
Now the function $\lambda$ is real on this curve and therefore $a$ must be real.
In fact $1\le\lambda<\infty$ on this semi-circle and so $1<a<\infty$ for
the critical point to lie between $\sigma=0$ and $\sigma=1$. 

Equation (\ref{betaG2}) is easily integrated with respect to $s$ to give
\begin{equation}
\label{exps}
C\hbox{e}^{-s}=\left( {\lambda -1\over a-1} \right)^{a-1} \left({a\over\lambda}\right)^a,
\end{equation}
where $C$ is an arbitrary integration constant (complex in general). 
Since the function $s(v)$ is assumed to be real along any flow line the complex
phase of the right hand side of the above equation must be constant and equal to
the phase of $C$. Thus the trajectory is determined by the phase of $C$ and
we can easily plot the integral curves of (\ref{betaG2}) by doing a contour plot
of the complex phase of the right hand side of (\ref{exps}). Two such plots are shown in figure 2,
the first is for the $\GU$ case, $a=2$, and has already been presented in \cite{BDb},
while the second corresponds to the choice $a=1.0749$. The distortion of the flow
away from the ideal $\GU$ flow is manifest in the lower
figure (Khmel'nitskii's original analysis, \cite{Khmel}, is only compatible with $\GU$).
One of the main points of this paper is to suggest that such a distortion
is an inevitable consequence of having Landau levels that are not completely
spin split. Some such distortion is a likely consequence of any flow, including
those that are not meromorphic. 

Although the quantitative shape of the flow shown in figure 2 does not depend
on the unknown function $s(v)$, provided only that it is monotonic,
this function must still be specified in order
to make further progress. In order to pin this function down, at least for Hall-Hall
crossovers, consider the semi-circular arches on which $\lambda$ is real
and greater than unity. We can always choose $s=0$ at the critical point
where $\lambda=a$, so $C=1$ in (\ref{exps}). Now note that the flow
diagrams in figure 2 are completely unaffected by a rescaling of $s(v)$. It is 
$v={ne\over T^\mu}\Delta\nu$ that is the physical parameter, so consider $v(s)$.
A rescaling of $s$ should not affect the flow diagram therefore $v(s)$ must be a scale
invariant function of $s$ and the only such function is a power law $v\propto s^{1/\alpha}$,
and so one concludes that $s=({A\Delta\nu\over T^\mu})^\alpha\propto v^\alpha$, for some
real constants $\alpha$ and $A$.\footnote{A different argument for this power law dependence
is given in \cite{BDb}.} This relates to the earlier observation that the $\beta$-functions
(\ref{analyticbeta}) have a pole at the critical values of $\sigma_c$ where $\lambda(\sigma_c)=a$.
Na{\rm\"\i}vely this looks disastrous --- a critical point should be associated with a zero
of the $\beta$-function not a singularity! Indeed experimentally $\sigma_{xy}$ and
$\sigma_{xx}$ are perfectly finite and smooth as the magnetic field passes through
$B_c$ (for finite $T$), 
and ${d\sigma\over dv}$ is finite at $v=0$. However, as emphasised previously, 
these are not Wilsonian $\beta$-functions and the resolution of this problem lies in the 
exponent $\alpha$. With $\beta={d\sigma\over ds}$ and $s\propto v^\alpha$ we have
\begin{equation}
\beta={d\sigma\over ds}\propto {1\over v^{\alpha-1}}{d\sigma\over dv},
\end{equation}
so, with ${d\sigma\over dv}$ finite at $v=0$, $\beta$ will diverge if $\alpha>1$.
Indeed a careful analysis of the analytic form of $\beta$ in equation (\ref{analyticbeta})
shows that $\beta\approx{1\over \sqrt{s}}$ near the critical point $\sigma=(1+i)/2$
and $\alpha=2$. This is also true for the spin degenerate form (\ref{betaG2}).

The explicit analytic form of $\sigma(A\Delta\nu/T^\mu)$, within the framework of
the meromorphic ansatz (\ref{betaG2})
can now be obtained using the same methods as in \cite{BDb}.
First express the invariant function $\lambda(\sigma)={\vartheta_2^4\over \vartheta_3^4}$
in terms of elliptic integrals of the second kind
\begin{equation}
K(k)=\int_0^{\pi/2}{d\varphi\over\sqrt{1-k^2\sin^2\varphi}},
\end{equation}
using the classical formulae
\cite{WW}
\begin{eqnarray}
\vartheta_2=\sqrt{2kK(k)\over\pi},\qquad
\vartheta_3=\sqrt{2K(k)\over\pi},\qquad
\vartheta_4=\sqrt{2k^\prime K(k)\over\pi},
\end{eqnarray}
(where the modulus $k$ is related to $\sigma$ by 
$\hbox{e}^{-{\pi K^\prime\over K}}=\hbox{e}^{i\pi\sigma}$
and $K^\prime(k)=K(k^\prime)$, with $(k^\prime)^2:=1-k^2$ the complementary modulus).
This gives $\lambda=k^2$ and equation
(\ref{exps}) reads, with $C=1$,
\begin{equation}
\hbox{e}^{-s}=\left({k^2-1 \over a-1}\right)^{a-1}\left({a \over k^2}\right)^a
\end{equation}
which, for a fixed $a$, defines $k^2$ in terms of $s$.
For $\GU$, with $a=2$, this is easily solved to give $k(s)$ explicitly,
but for a more general value of $a$ the relation must remain implicit.

Since $1<k^2<\infty$ on the semi-circle spanning $\sigma=0$ and $\sigma=1$
the elliptic integrals for the crossover are complex when written in  terms of $k$,
which makes it difficult to pick out the real and imaginary parts
of $\sigma$ from the relation $\hbox{e}^{-{\pi K^\prime\over K}}=\hbox{e}^{i\pi\sigma}$.
So, as in \cite{BDb}, this relation is re-expressed in terms of $w=1/k$, so
that $0<w<1$ and $K(w)$ and $K^\prime(w):=K(w^\prime)$ are then real,
\begin{equation}
\label{analyticba}
\sigma(s)={K^\prime(w)\{K^\prime(w)+iK(w)\}\over [\{K(w)\}^2 + \{K^\prime(w)\}^2]}.
\end{equation}
This equation, together with the implicit definition of $w(s)$
\begin{equation}
\label{analyticbb}
\hbox{e}^{-s}=\left({w^{-2}-1 \over a-1}\right)^{a-1}\bigl({a w^2}\bigr)^a,
\end{equation}
and the relation derived earlier for $s$ on the semi-circle
\begin{equation}
\label{sdef}
s=\left({A\Delta\nu\over T^\mu}\right)^2,
\end{equation}
completely determine the crossover conductivities as functions of the scaling
variable ${\Delta\nu\over T^\mu}$ in terms of two unknown, sample specific,
constants, $A$ and $a$. The exponent $\mu$ is believed to be universal
and has been measured to be $\mu=0.45\pm0.05$, \cite{Shaharetal}.
The conductivities, and corresponding resistivities, for the integer
transition $\nu: 1\rightarrow 2$ in the $\GU$ case ($a=2$)were first derived in \cite{BDb}
and are reproduced here in figure 3, with the choice $A=40$
and $\mu=0.5$, as functions of $\Delta\nu$ at the four different temperatures
used in \cite{Shaharetal} for this transition.

Before comparing these analytic expressions with the experimental data we
must address the issue of how the experimental longitudinal conductivity is
related to its theoretical cousin.

\section{Renormalisation of $\rho_{xx}$}

Experimentally the longitudinal conductivity cannot be measured directly.
What is actually measured is the longitudinal and transverse resistances
from which are inferred the longitudinal and transverse resistivities.
In two dimensions, of course, resistivities and resistances have the same
dimensions but they can still differ by dimensionless geometric factors.
The transverse resistivity and the transverse resistance are identical,
independent of any such geometrical factors --- a fact which is essential
for the accuracy of the Hall quantization --- but this is not the
case for the longitudinal resistivity, $\rho_{xx}$ and resistance, $R_{xx}$.
When experimental resistivities are quoted it is actually $R_{xx}$ that is
measured, together with the length $L$ and width $W$ of a rectangular sample,
and the formula
\begin{equation}
\label{rhoexp}
R_{xx}={L\over W}\rho^{exp}_{xx}
\end{equation}
is used to determine the experimental longitudinal resistivity, 
$\rho^{exp}_{xx}$.
As fabrication techniques improve $W$ and $L$ become more accurately determined
and so $\rho^{exp}_{xx}$ becomes more accurately measured. 
However let us examine equation (\ref{rhoexp}) carefully and ask what assumptions
go into the derivation and how $\rho^{exp}$ might be related to the $\rho_{xx}$
that a theoretician might quote. The derivation of (\ref{rhoexp}) is
extremely simple because the assumptions are grossly over simplified, \cite{Cage}.
In the absence of any magnetic field consider a 
rectangular slab carrying current density $j=\sigma_{xx}{V\over L}$ running parallel to
its long axis, where $V$ is the longitudinal voltage driving the current. 
Using $0<y<W$ as a Cartesian co-ordinate along the
width the total current is
\begin{equation}
I=\int_0^W jdy =\sigma_{xx}{V\over L}W={V\over R_{xx}}
\end{equation}
so $\rho_{xx}=1/\sigma_{xx}={W\over L}R_{xx}$, which is equation (\ref{rhoexp}).
Some important assumptions that go into this derivation are:
\begin{itemize}
\item{A perfectly rectangular sample.}
\item{Electron trajectories which are classical, linear and parallel.}
\item{No perturbations due to source and drain contacts or Hall voltage probes.}
\item{Neglect of any magnetic fields generated by the currents themselves.}
\item{Electrons scatter before reaching the drain.}
\end{itemize}
As fabrication techniques improve the first assumption may be getting better
and better but the others are somewhat questionable.
Impurities or electron-electron interactions are liable to make electron
trajectories deviate from straight lines and the last assumption may break
down for very pure samples at low temperatures, for example.
A more correct relation between longitudinal resistance and resistivity would read:
\begin{equation}
\label{rhoth}
R_{xx}={L\over W}f(W/L)\rho_{xx}
\end{equation}
where $f(W/L)$ is some function of the geometry that tends to unity
as $W/L\rightarrow 0$, and this is all one can really say.
A relation similar to (\ref{rhoth})
has been studied when an external magnetic field is present in \cite{Rendell+Girvin}, however
this analysis suffers from the same difficulties as listed above, and indeed
some of these are discussed in that reference.

If $W/L$ is not small we should expect differences between
the experimentalist's $\rho^{exp}_{xx}$ in equation (\ref{rhoexp}) and the theoretician's
$\rho_{xx}$ in equation (\ref{rhoth}). 
If there is such a renormalisation of $\rho_{xx}$ how can this be reconciled
with the experimentally observed semi-circle law, \cite{Hilkeetal}? Would this not
be modified if the experiment had used the wrong normalisation for $\rho_{xx}$?
The answer is yes, in general, but no for Laughlin-insulator transitions only ---
and in fact
the best experimental data for the semi-circle law come from the Laughlin-insulator
transitions, \cite{Hilkeetal}. 
Rescaling $\rho_{xx}$ for the Laughlin-insulator transitions
($\sigma:{1\over q}\rightarrow 0$, with integral $q$) 
does not affect the semi-circle law because these transitions are described by the semi-circle 
$\sigma={q+iw\over q^2+w^2}$, $0<w<\infty$, which is mapped to a vertical line 
$\rho=-q+iw$ in the $\rho$-plane by $\sigma=-1/\rho$. Rescaling $\rho_{xx}=\Im\rho$
does not change the nature of these vertical lines
and so merely serves to re-parameterise the semi-circle in the $\sigma$-plane, 
but it is still a semi-circle.
This is not true for Hall-Hall transitions, such as $\sigma:2\rightarrow 1$
for example.
The semi-circle $\sigma={2+w^2+iw\over 1+w^2}$ is mapped to $\rho={-(2+w^2)+iw \over 4+w^2}$
which is also a semi-circle. Rescaling $\rho_{xx}$ would distort this semi-circle into
an ellipse in the $\rho$-plane which would then 
correspond to an ellipse in the $\sigma$-plane. 

Indeed this suggests an alternative technique for determining $\rho_{xx}$ from
the experimental data, which does not rely 
on (\ref{rhoexp}). If one assumes the Law of Corresponding States,
including the particle-hole transformation, to
be a symmetry at low temperatures (possibly a more plausible assumption than classical,
linear trajectories for electrons) then the semi-circle law follows \cite{BDD}. For a Hall-Hall
transition it has just been shown that there is at most one normalisation of $\rho_{xx}$
that leads to a semi-circular transition --- so if we {\it assume} a semi-circular transition
we can use this to determine the normalisation of $\rho_{xx}$ for a Hall-Hall transition. 
This provides a technique
for determining $\rho_{xx}$ from the resistance $R_{xx}$ which is completely
independent of any geometrical factors or assumptions --- the only assumption is
that the Law of Corresponding states is a symmetry at low enough temperatures,
which is well motivated \cite{KLZ}. This will not work for Laughlin state-insulator
transitions since these correspond to straight vertical lines in the $\rho$-plane, which
are always semi-circles in the $\sigma$-plane regardless of the normalisation of $\rho_{xx}$,
but this very fact can be turned to our advantage. Any deviation from a semi-circular arch
in the $\sigma$-plane for a Laughlin state-insulator transition
can be interpreted as a sign that the temperature is not yet 
low enough for the Law of Corresponding States to be a valid symmetry and this gives
a quantitative measure of when the temperature is ``low enough'' for the normalisation
of $\rho_{xx}$ to be fixed for a Hall-Hall transition.

In the following section we shall compare the analytic predictions of the previous
section with the available experimental data, allowing for the possible ambiguity
in the normalisation of the experimentally determined value of
$\rho^{exp}_{xx}$ in Laughlin-insulator transitions discussed above.

\section{Comparison with experimental data}

The analytic prediction of section 4 for $\GU$, based on the assumption of meromorphicity
and derived in \cite{BDb},
is reproduced for the transition $\nu:2\rightarrow 1$
in figure 3. The values of $A$ and $\mu$, as well as the values
of the temperature, used here were specially chosen
to give a good fit to the data of \cite{Shaharetal} and the analytic curves are 
displayed in conjunction with the experimental data in figure 4. The agreement
appears to be quite good without any rescaling of $\rho_{xx}$, 
despite the lack of any physical justification for
the meromorphic ansatz. It is to be stressed that there is really only {\it one}
adjustable parameter in this fit, the value of the constant $A$ which is
just a choice of scale for the horizontal axis. The exponent $\mu$ is not
independent, but is measured in this experiment to be $\mu=0.45\pm 0.05$ so there is only a
small leeway available to vary it to fit the curves. The fit is not perfect
but it seems remarkable that it is so good with only one parameter at our 
disposal!
For the Laughlin-insulator transition $\nu:1\rightarrow 0$ the theoretical
prediction is that the critical resistivity is $\rho^c_{xx}=1$ while the
experimental data indicate a value some 30\% larger than this. So we use
the freedom to re-scale $\rho^{exp}_{xx}$ discussed in the last section
to argue that the true normalisation for $\rho_{xx}$ is such that the
critical point {\it should} be at unity and re-scale the theoretical $\rho_{xx}$
by $1.3$ to obtain figure 5, which is super-imposed on the experimental
data in figure 6. Again the temperatures are chosen to agree with those
used in the actual experiment, but this time the value $A=60$ is found to
give the best fit (there is no reason to expect $A$ to be universal).
The same sample was used in these two experiments, so the interpretation
suggested here requires a different normalisation for $\rho_{xx}$ in the
two transitions. This could be accommodated, for example, by postulating
a $B$ dependence in the geometrical function $f$ of equation (\ref{rhoth}),
causing $f$ to vary by 30\% as $B$ is increased by a factor of three,
which is the ratio of the critical fields in the two transitions,
but with $f$ being fairly constant under variations of $B$ of order 10\%,
which is the range depicted in the experimental transitions.  

A second experiment, with excellent support for a semi-circular transition,
is that of Hilke et al., \cite{Hilkeetal}. In this experiment the critical point in the
$1\rightarrow 0$ transition is most certainly not at $\sigma={1+i\over 2}$.
Two possible explanations of this are are: i) the Landau levels
come in closely related pairs, so that the symmetry group is $\G2$ with $a\ne 2$ and 
not $\GU$; ii) the group is $\GU$ but 
the experimental $\rho^{exp}_{xx}$ differs from the theoretical one
by a factor of $1.7$. Using the predictions of the meromorphic ansatz
we can test these two hypothesis and by far the most successful is the latter ---
it is simply impossible to get a good fit with the former assumption while
the theoretical curves for the latter, after increasing  $\rho_{xx}$ by 1.7,
are shown in figure 7 and compared to the experimental data in figure 8.
The data here are plotted against $B$ rather than $\nu$ and converting
$\nu$ to $B$ requires knowing the charge carrier density. In terms of the magnetic field,
$B\propto 1/\nu$, equation (\ref{sdef}) gives the relation between
$B$ and the scaling variable $s$
\begin{equation}
\label{Cdef}
{\Delta B\over B_c}=\mp{A^\prime T^\mu \sqrt{s}\over 1 \pm A^\prime T^\mu\sqrt{s}}
\end{equation}
where $A^\prime$ is related to the constant $A$ in equation (\ref{sdef}) 
by $A^\prime\nu_c=1/A$, with $\nu_c$ the
critical filling factor.
The best value of $A^\prime$ to fit the data of \cite{Hilkeetal} is 
$0.05$ while the optimal value of $\mu$ is $0.29$ which is certainly
a little low for comfort, but is just compatible with three standard deviations
of the value in \cite{Shaharetal}
(no experimental value for $\mu$ is  quoted in the paper \cite{Hilkeetal},
so we work with the value given in \cite{Shaharetal}, because this is believed to
be universal). 
The highest temperature at $3.2$K is a very
poor fit, but all the experimental evidence is that this temperature is
too high for $\GU$ symmetry to be valid --- there is strong deviation in
the experimental data away from
the semi-circle law at this temperature too.
So again the agreement between the experimental data and the theoretical predictions
of the meromorphic ansatz is very good, provided one re-scales $\rho^{exp}_{xx}$
and assumes $\GU$ symmetry. The actual values of $L$ and $W$ in this sample are
$50\mu$m and $600\mu$m so $W/L=0.08$ is small but not really close to zero. 

 Of course just because one analytic suggestion appears to give a good fit to
the experimental data does not mean that it has to be the correct one.
The experimental data for the transition $\sigma:1\rightarrow 0$ can also
be fitted by an exponential, \cite{SHLTSR}, but when this is done it appears
to indicate that scaling, and therefore super-universality, may be in trouble.
The authors of \cite{SHLTSR} fit the form 
$\rho_{xx}(\Delta\nu)={e}^{-{\Delta\nu\over F(T)}}$
to their experimental data, where $F(T)$ is a function of $T$ to be determined.
They find that $F(T)\propto T^\mu$ is incompatible with the data
for any $\mu$ and suggest instead a linear form,
$F(T)=\tilde\alpha T+\tilde\beta$
with $\tilde\alpha$ and $\tilde\beta$
non-zero constants. Since $\tilde\beta\ne0$ gives the best fit 
this is interpreted as a violation of the scaling hypothesis. 
It was suggested in \cite{BDb} that it may be possible to recover scaling with
a different functional fit --- in particular the form suggested here
in terms of elliptic integrals seems promising --- but a definitive answer
would require a quantitative analysis, such as a $\chi^2$-fit, which
would require access to the actual experimental numbers rather than just the figures
in \cite{SHLTSR}. An important difference between the exponential and the elliptic
integral forms is that, near the Hall plateau at which $\Delta\nu>0$, 
$\sigma_{xx}\sim e^{-\Delta\nu}$ for the former while 
$\sigma_{xx}\sim 1/(\Delta\nu)^2$ in the latter. It is difficult to distinguish
visually between  these two possibilities with the current data.
 
 Returning to the difference between $\GU$ and $\G2$ all the experimental data
considered here are well described with $\GU$.
It would be very interesting to see good experimental crossover curves at different
temperatures for samples with very small spin splitting, at temperatures low
enough for the semi-circle law to be valid over a range of temperatures. 
The arguments given in section two would
imply that it may be $\G2$ that is relevant then rather than $\GU$ and 
the sample dependent
parameter $a$, which parameterizes the position of the critical point, would
have to be taken from experiment rather than coming as a prediction from the
symmetry.

\section {Conclusions}

It has been argued that the rich duality structure of the quantum Hall effect, as
embodied in the infinite order discrete non-abelian group $\GU$, which is the
mathematical description of the Law of Corresponding States, should be modified when
the Landau levels are organized into close pairs, for example
when the spin splitting is small relative to the cyclotron energy. This modification
requires replacing $\GU$ with the smaller, but nevertheless still infinite order,
group $\G2$. In terms of the Law of Corresponding states this replacement is equivalent
to replacing the Landau level addition transformation, $\L:\sigma\rightarrow\sigma+1$, with
$\sigma\rightarrow\sigma+2$.
The derivation of the semi-circle law from duality symmetry is still valid 
if one assumes that the particle-hole transformation of the spin split case,
$\Ph:\sigma\rightarrow 1-\bar\sigma$, is replaced 
with $\sigma\rightarrow 2-\bar\sigma$.
The main difference between the groups $\GU$ and $\G2$
is that for the later the critical point in the crossover between two
Hall plateaux (or between a Laughlin state and the  insulating phase) is no longer fixed
by duality, as it is in $\GU$ case, but can be anywhere on the semi-circle
spanning the two states in the complex $\sigma$-plane. This results in
a distortion of the flow diagram shown in the lower diagram of figure 2.

The experimental data often show a critical point in the crossover which
is not at the point predicted by $\GU$ symmetry, i.e. at $\sigma_c={1+i\over 2}$
for the Hall-Hall transition $\sigma:2\rightarrow 1$. Two possible causes of this are
investigated here: i) it is due to the breaking of $\GU$ to $\G2$ or ii) it is
due to a renormalisation of the experimentally determined longitudinal resistivities
so that they do not coincide with their theoretical values. Without any further assumptions
one cannot distinguish between these two possibilities. 

With the added assumption that the $\beta$-function described in section three are meromorphic
functions of the complex conductivity an analytic form is obtained for 
$\sigma_{xx}({\Delta B\over T^\mu})$ and  $\sigma_{xy}({\Delta B\over T^\mu})$.
For the $\G2$ case there are two free  parameters in the resulting
expression, equations (\ref{analyticba})-(\ref{sdef}), apart from the exponent $\mu$ ---
these correspond to the normalisation of the horizontal axis in figure 3 and the
angular position of the critical point in the crossover. For integer transitions
the critical point lies at the top of the semi-circle when the larger symmetry $\GU$
is relevant but can be anywhere on the arch for the smaller group $\G2$.

However a simplified analysis
of the relation $\rho_{xx}={W\over L}R_{xx}$, used to determine the longitudinal resistivity
from the experimentally measured longitudinal resistance, shows that it should only
be considered to yield the correct normalisation for $\rho_{xx}$ in the limit 
${L\over W}\rightarrow\infty$. It is suggested that a more reliable technique for
determining the correct normalisation is to ascertain the temperature range
for which the Laughlin-insulator transition is semi-circular and to
assume that the semi-circle law is also valid for Hall-Hall transitions in this regime.
This would then fix a unique value for the ratio  $\rho_{xx}/R_{xx}$ for the Hall-Hall
transitions.

A comparison with some of the available experimental data, 
gives very good fits (figures 4, 6, and 8) for $\GU$ duality --- provided one 
renormalises the experimental value of $\rho_{xx}$, at least for the Laughlin-insulator
transitions. This author is not aware of any experimental data in support of the
breaking of $\GU$ to $\G2$ but, if the ideas presented here are correct,
it ought to be possible to see this breaking in samples where the spin splitting
is small and the temperature is low enough for the semi-circle law to be valid.

The main difficulty with the present analysis is of course the ad hoc assumption
of meromorphic $\beta$-functions. There is no theoretical justification for this
assumption, either microscopic or mesoscopic, but in view of the remarkable agreement
with experiment that it produces (figures 4, 6, and 8) it would seem to merit
some consideration. The assumption of complex analyticity is a very natural one in
the context of super-symmetric gauge theories, where the groups $\GU$ and
$\G2$ also manifest themselves as duality symmetries, but this is a consequence
of super-symmetry in these models. While it is not impossible that super-symmetry may be
relevant to the quantum Hall effect it is certainly not established and probably not
accepted by most workers in the field. The original motivation for testing an analytic
ansatz was simply mathematical elegance and it is certainly a surprise that it
seems to give answers that agree so well with experiment. Nevertheless it seems
worthwhile pursuing this hypothesis in the light of said agreement and a possible
physical justification for complex analyticity of the $\beta$-functions is currently
under investigation.  

It is a pleasure to thank the Theoretical Physics group, Institute for Nuclear Physics, 
Orsay, where most of this work was carried out, for their hospitality and in particular
Yvon Georgelin for useful discussions.
I would also like to thank Cliff Burgess for helpful discussions on crossover
in the quantum Hall effect.

\pagebreak
\pagebreak

\pagebreak
\centerline {\bf FIGURE CAPTIONS}
\bigskip

\vbox{\noindent{\it Figure 1: Landau levels and spin splitting.} 

\noindent For spinless
electrons (centre) the Landau levels are evenly spaced; for large spin splitting
(left) it is still reasonable to assume that the physics of each Landau level is independent
of how many lower levels are full, provided the splitting is not so large
that some levels begin to get close to each other; 
for small spin splitting (right) the physics of a pair
of levels is independent of how many lower pairs are filled but the physics of electrons in
the upper member of a pair could well be influenced by electrons in the lower member.}
\bigskip

\vbox{\noindent{\it Figure 2: Flow diagrams predicted by the Law of Corresponding States
and modular symmetry.} 

\noindent The upper diagram is for $\GU$ symmetry and the lower diagram
is for $\G2$ symmetry --- the latter is expected to be 
associated with small spin splitting. Although the detailed form of these
flows relies on the meromorphic ansatz for the $\beta$-functions described in the
text, the general topology is independent of this assumption and the same topology
of flows would result also from non-meromorphic $\beta$-functions.}
\bigskip

\vbox{\noindent{\it Figure 3: Analytic crossover for the $2\rightarrow 1$ transition as a function
of $\Delta\nu$.} 
\noindent The analytic form of the crossover for the Hall-Hall
transition $\sigma:2\rightarrow 1$ for $\GU$ symmetry. The horizontal scale
is set by $A=40$, where $A$ is defined in (\ref{sdef}), and the exponent $\mu$ is 
$0.50$ --- the four curves are
for $T=42,70,101$ and $137$mK (these are the temperatures used in the
experiment described in \cite{Shaharetal}).}
\bigskip

\vbox{\noindent{\it Figure 4: Comparison of theoretical predictions to experimental data
for the Hall-Hall transition $2\rightarrow 1$.} 

\noindent The theoretical curves of figure 3 are compared with the experimental 
curves of \cite{Shaharetal}.}
\bigskip

\vbox{\noindent{\it Figure 5: Analytic crossover for the $1\rightarrow 0$ (insulator) transition
as a function of $\Delta\nu$.} 

\noindent The analytic form of the crossover for the Hall-insulator
transition $\sigma:1\rightarrow 0$ for $\GU$ symmetry. The horizontal scale
is set by $A=60$, the exponent $\mu$ is $0.50$ and the four curves are
for $T=42,84,106$ and $145$mK. The longitudinal resistivity, $\rho_{xx}$, has
been rescaled by a factor of $1.3$ in order to agree with the experimental data
of \cite{Shaharetal}, as described in the text.}
\bigskip

\vbox{\noindent{\it Figure 6: Comparison of theoretical predictions to experimental data
for the Hall-insulator transition $1\rightarrow 0$.} 

\noindent The theoretical curves of figure 5 are compared with the experimental curves 
of \cite{Shaharetal}.}
\bigskip

\vbox{\noindent{\it Figure 7: Analytic crossover for the $1\rightarrow 0$ (insulator) transition
as a function of $\Delta B/B_c$.} 

\noindent The analytic form of the crossover for the Hall-insulator
transition $\sigma:1\rightarrow 0$ for $\GU$ symmetry. The horizontal scale
is set by $A^\prime=0.05$, where $A^\prime$ is defined in equation (\ref{Cdef}), and
the exponent $\mu$ is $0.29$ --- the five curves are
for $T=0.5,0.8,1.2,1.8$ and $3.2$K which are the temperatures used in the experiment described
in \cite{Hilkeetal}. The longitudinal resistivity, $\rho_{xx}$, has
been rescaled by a factor of $1.7$ in order to agree with the experimental data
of \cite{Hilkeetal}, as described in the text.}
\bigskip

\vbox{\noindent{\it Figure 8: Comparison of theoretical predictions to experimental data
for the Hall-insulator transition $1\rightarrow 0$.} 

\noindent The theoretical curves of figure 7 are compared with the experimental curves of \cite{Hilkeetal}.
The highest temperature, $T=3.2$K, is obviously a poor fit to the data
but it is clearly stated in the experimental paper that this temperature violates
the semi-circle law and so this is too high for $\GU$ symmetry to be applicable.
The inset shows the experimental verification of the semi-circle law with the critical
point marked at $B_c$. This point can be moved to the top of the arch by dividing the
experimental values for $\rho_{xx}$ by $1.7$.}

\pagebreak
\centerline{ }
\vskip 1cm
\vtop{
\includegraphics{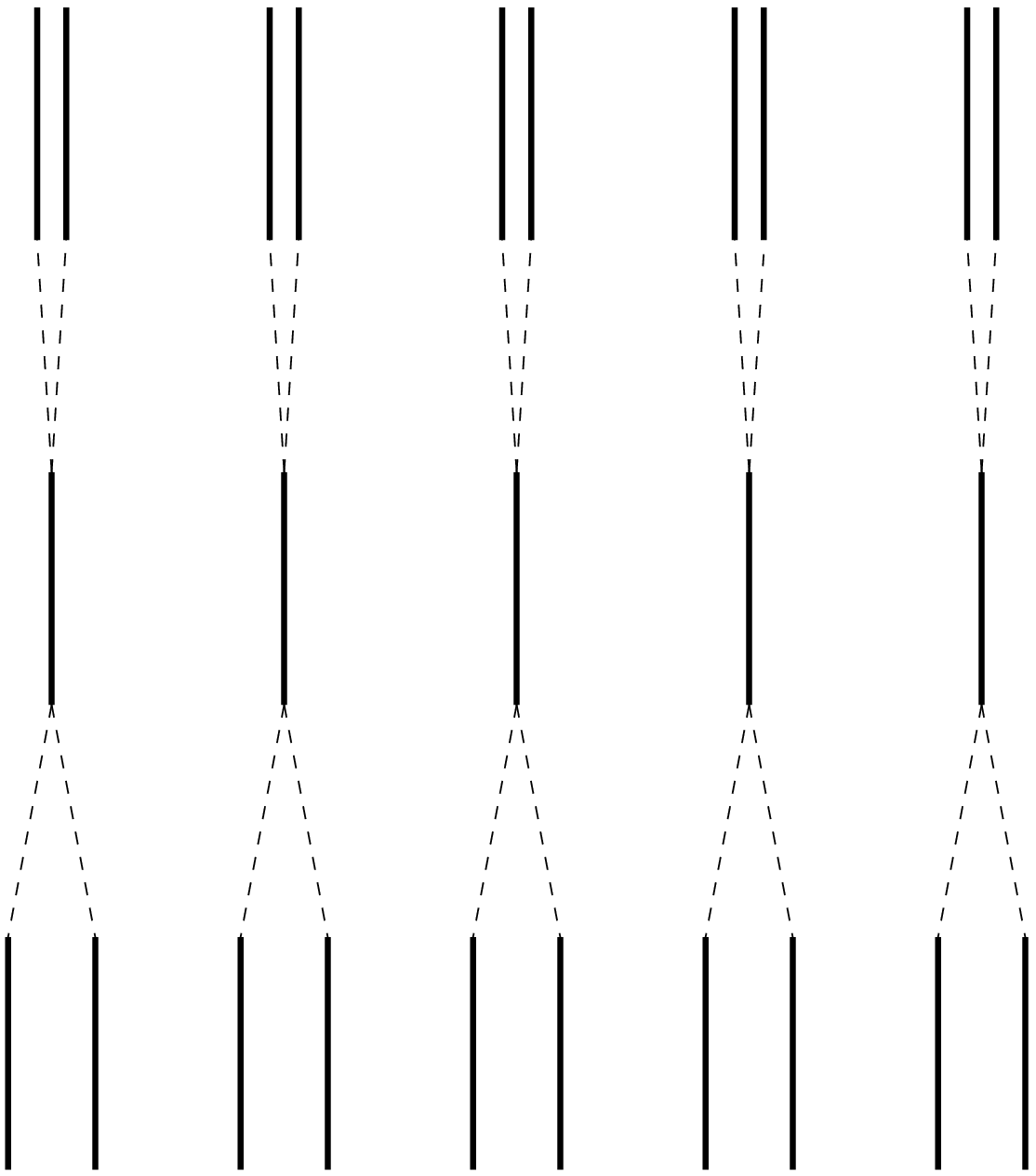}
\vskip12cm
\centerline{\bf Figure 1}
}

\pagebreak
\vtop{
\centerline{}
\includegraphics{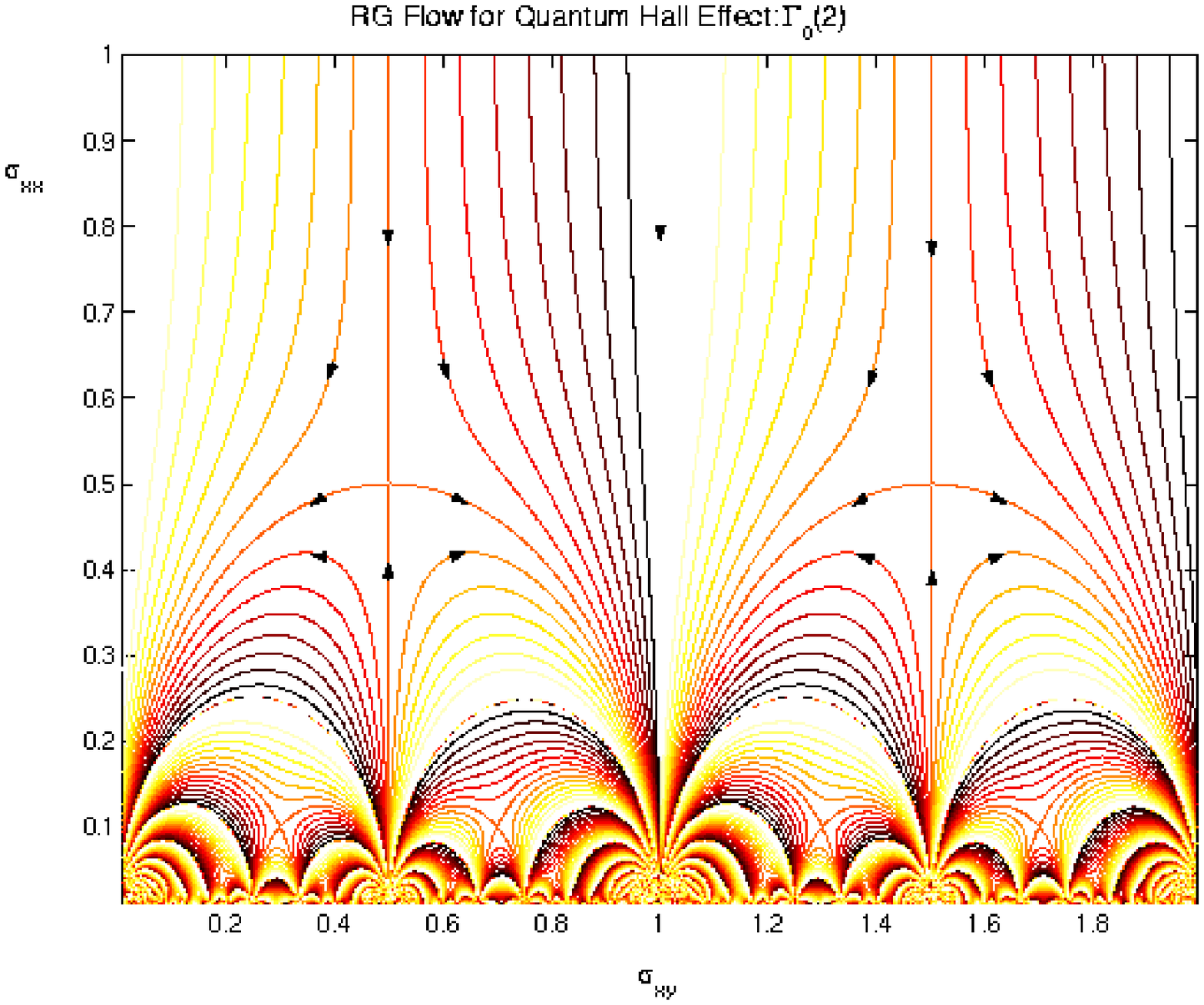}
\includegraphics{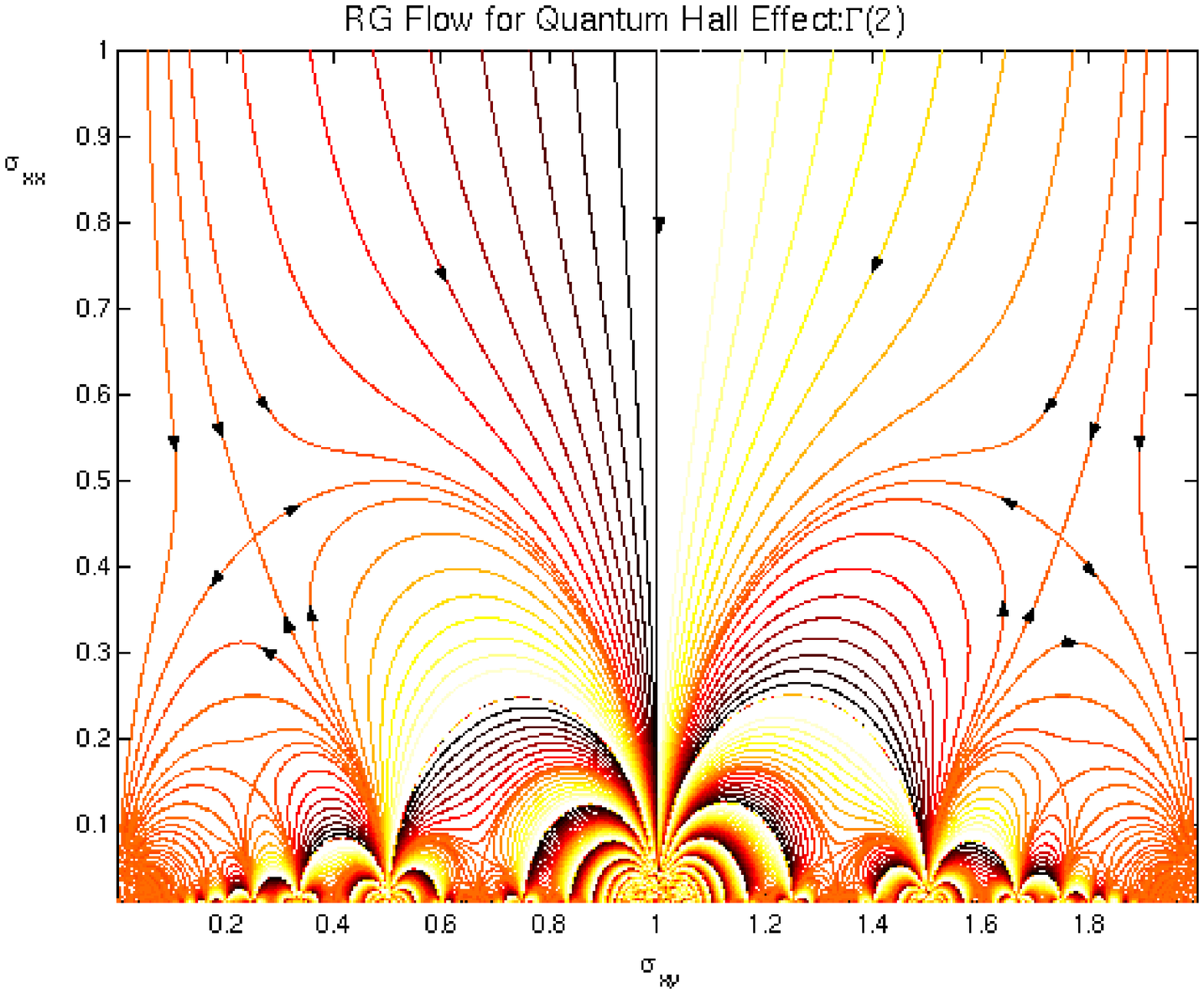}
\vskip 16.5cm
\centerline{\bf Figure 2}
}

\vtop{
\centerline{ }
\hskip -3cm
\includegraphics{fig3a.ps}
\vskip -0.1cm
\hskip 1cm $\rho_{xy}$
\vskip 4cm
\hskip 1cm $\rho_{xx}$
\vskip 1.5cm
\centerline{$\Delta\nu$}
\hskip -3cm
\includegraphics{fig3b.ps}
\vskip 1.5cm
\hskip 1cm $\sigma_{xy}$
\vskip 5cm
\hskip 1cm $\sigma_{xx}$
\vskip 1.8cm
\centerline{$\Delta\nu$}
\centerline{\bf Figure 3}
}    

\pagebreak
\vtop{
\centerline{}
\hskip -3cm
\includegraphics{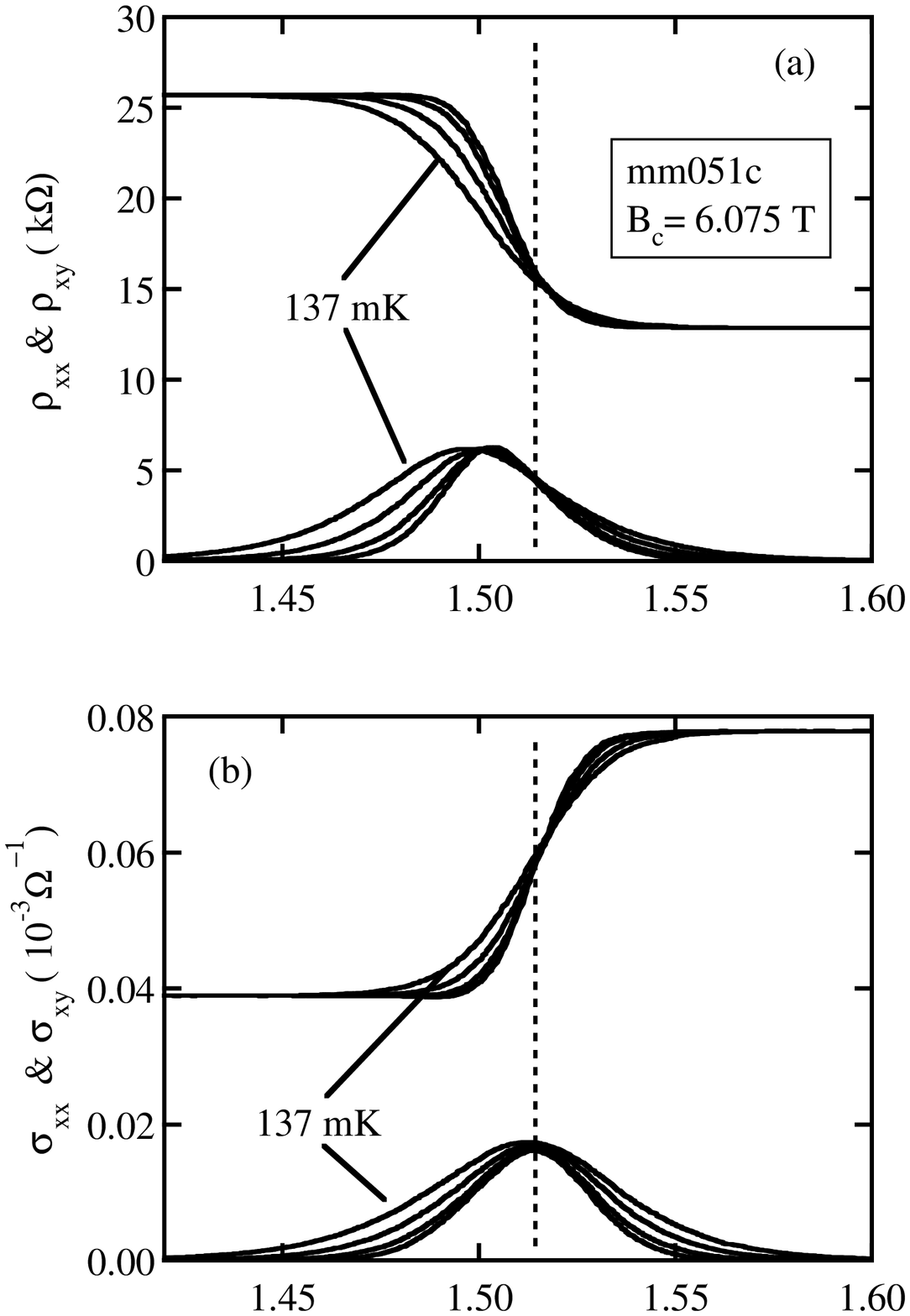}
\includegraphics{fig3a.ps}
\includegraphics{fig3b.ps}
\vskip 18.4cm
\centerline{\bf Figure 4}
}

\pagebreak
\centerline{ }
\vtop{
\includegraphics{fig5.ps}
\centerline{ }
\vskip 2.2cm
\hskip 1.5cm $\rho_{xx}$
\vskip 3.3cm
\hskip 1.5cm $\rho_{xy}$
\vskip 3cm
\centerline{$\Delta\nu$}
\vskip 1cm
\centerline{\bf Figure 5}
}

\pagebreak
\centerline{ }
\vskip -10cm
\hskip -2.5cm
\vtop{
\includegraphics{fig6.ps}
\includegraphics{fig5.ps}
}
\vskip 20cm
\centerline{\bf Figure 6}
\pagebreak

\vtop{\includegraphics{fig7.ps}
\vskip 0.2cm
\hskip 11cm $\nu:0\rightarrow 1$
\vskip 1cm
\hskip -0.5cm $\rho_{xx}$
\vskip 4.3cm
\hskip -0.5cm $\rho_{xy}$
\vskip 1.7cm
\hskip 8.7cm $\Delta B/B_{crit}$
\vskip 1cm
\centerline{\bf Figure 7}
}

\pagebreak
\hskip -3cm
\vtop{
\includegraphics{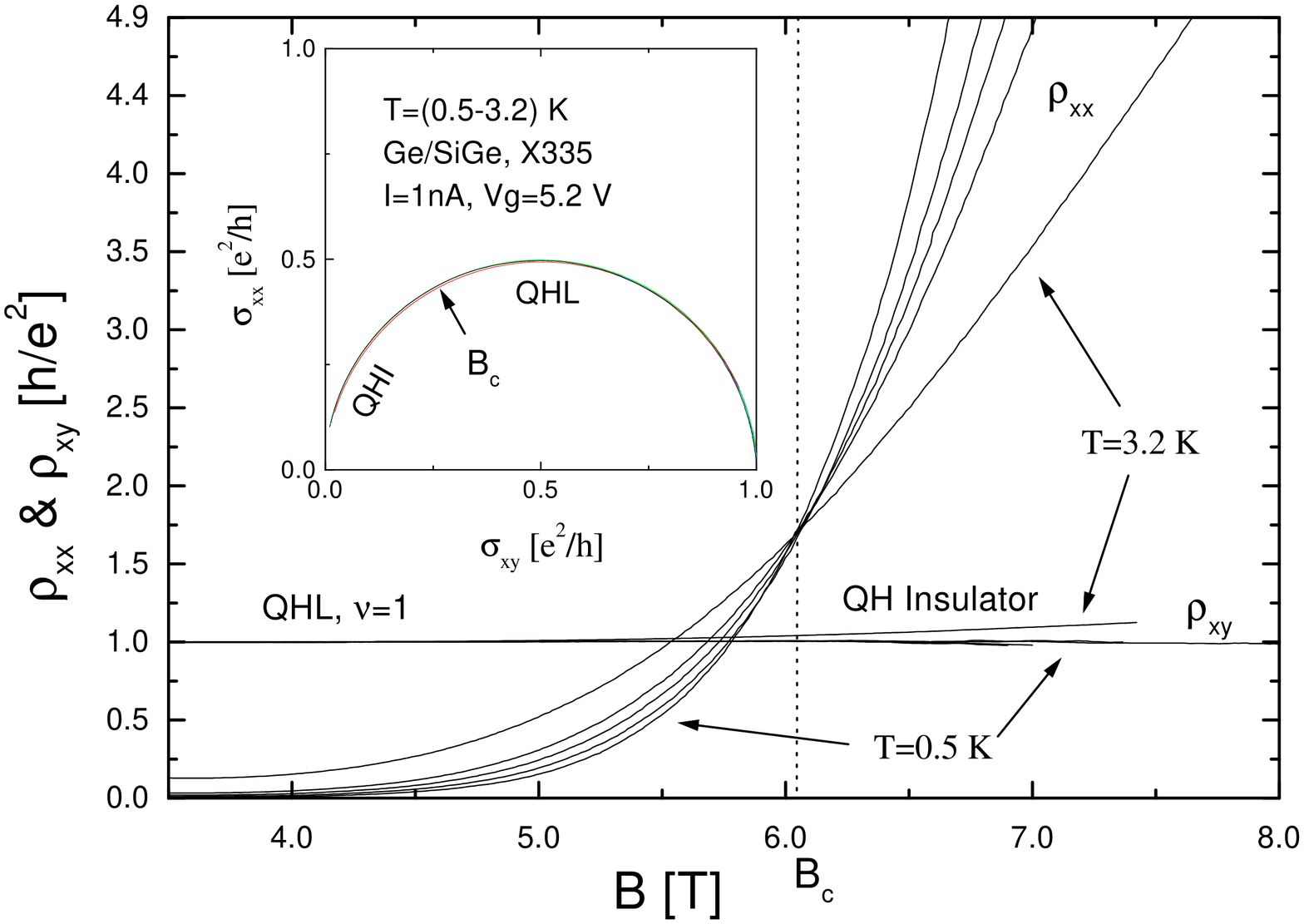}
\includegraphics{fig7.ps}
\vskip 10cm
\hskip 7.5cm{\bf Figure 8}
}

\begin{thebibliography}{99}
\bibitem{Jain}J.K.~Jain, Phys. Rev. Lett. {\bf 63}, 199 (1989)\hfill\break
J.K.~Jain, Adv. Phys. {\bf 41}, 105 (1992)
\bibitem{ZHK} S.-C.~Zhang, T.H.~Hansson and S.~Kivelson, Phys. Rev. Lett. {\bf 62},
82 (1989)
\bibitem{KLZ} S.~Kivelson, D-H.~Lee and S-C.~Zhang, Phys. Rev. {\bf B46},
2223 (1992)
\bibitem{Lutken+RossA} C.A.~L\"utken and G.G.~Ross, Phys. Rev. {\bf B45}, 11837
(1992)
\bibitem{Lutken+RossB} C.A.~L\"utken and G.G.~Ross, Phys. Rev. {\bf B48}, 2500
(1993)
\bibitem{Rey+Zee} S.-J.~Rey and A.~Zee, Nucl. Phys. {\bf B352}, 897 (1991)
\bibitem{Cardy} J.L.~Cardy and E.~Rabinovici, Nuc. Phys. {\bf B205}, 1
(1982); \hfill\break
J. L. Cardy, Nuc. Phys. {\bf B205}, 17 (1982)
\bibitem{Bertoldi} G. Bertoldi, {\sl Potts model: Duality, Uniformization and the 
Seiberg-Witten modulus}, cond-mat/9911383
\bibitem{Lutkenb} C.A.~L\"utken, J. Phys. A: Math.  Gen. {\bf 26}, L811-L817 (1993)
\bibitem{DualityReview} E. Kiritsis, {\sl Supersymmetry and Duality in Field Theory and String Theory}, hep-ph/9911525
\bibitem{Burgess+Lutkena} C.P.~Burgess and C.A.~L\"utken, Nuc. Phys. {\bf B500}, 367 (1997)
\bibitem{BDa} B.P.~Dolan, J. Phys. {\bf A32}, (1999) L243 
(cond-mat/9805171)
\bibitem{BDb} B.P.~Dolan, Nuc. Phys. {\bf 460B [FS]}, (1999) 297 (cond-mat/9809294)
\bibitem{Taniguchi} N.~Taniguchi, {\sl Nonperturbative Renormalisation Group Function for 
Quantum Hall Plateau Transitions Imposed by Global Symmetries}, cond-mat/9810334
\bibitem{Burgess+Lutkenb} C.P.~Burgess and C.A.~L\"utken, Phys. Lett. {\bf B451}, 365
(1999) (hep-th/9812396)
\bibitem{GWT-Mb} Y.~Georgelin, T.~Masson and J-C.~Wallet, {\sl $\Gamma(2)$ Modular Symmetry,
Renormalization Group Flow and the Quantum Hall Effect}, cond-mat/9906193
\bibitem{BDD} C.P.~Burgess, R.~Dib and B.P.~Dolan, {\sl Derivation of the Semi-circle 
Law from the Law of Corresponding States}, cond-mat/9911476
\bibitem{SW} N.~Seiberg and E.~Witten, Nuc. Phys. {\bf B426}, 19 (1994) (hep-th/9407087)
\bibitem{Koblitz} N.~Koblitz, {\sl Introduction to Elliptic Curves and Modular Forms}, 2nd ed., 
(Springer, 1993).
\bibitem{Rankin} R.A.~Rankin, {\sl Modular Forms and Functions}, 
(Cambridge University Press, 1977).
\bibitem{Lutkena} C.A.~L\"utken, Nuc. Phys. {\bf B396}, 670 (1993)
\bibitem{GWT-Ma} Y.~Georgelin and J-C.~Wallet, Phys. Lett. {\bf A224}, 303
(1997);
Y.~Georgelin, T.~Masson and J-C.~Wallet, J. Phys. {\bf A30}, 5065 (1997)
\bibitem{Wilczek+Shapere} A.~Shapere and F.~Wilczek, Nuc. Phys {\bf B320}, 669 (1989)
\bibitem{Fogler} M.M.~Fogler and B.I.~Shklovskii, Phys. Rev. {\bf B52}, 17366 (1995)
\bibitem{Khmel} D.E.~Khmel'nitskii, Pis'ma Zh. Eksp. Teor. Fiz. {\bf 38}, 
454 (1983) (JETP Lett. {\bf 38}, 552 (1983))
\bibitem{Pruisken} A.M.M.~Pruisken, Phys. Rev. Lett. {\bf 61}, 1297 (1988)
\bibitem{Ruzin}  A.M.~Dykhne and I.M.~Ruzin, Phys. Rev. {\bf B50}, 2369 (1994); 
I.~Ruzin and S.~Feng, Phys. Rev. Lett. {\bf 74}, 154 (1995)
\bibitem{Hilkeetal} M.~Hilke et al., Euro. Phys. Lett. {\bf 46}, 775 (1999) (cond-mat/9810217)
\bibitem{STSCSS} D.~Shahar, D.C.~Tsui, M.~Shayegan, J.E.~Cunningham,
E.~Shimshoni and S.L.~Sondhi, Solid State Comm. {\bf 102} (1997) 817
(cond-mat/9607127);
D.~Shahar, D.C.~Tsui, M.~Shayegan, E.~Shimshoni and S.L.~Sondhi,  
Science {\bf 274}, (1996) 589 (cond-mat/9510113)
\bibitem{Ritz} A.~Ritz, Phys. Lett. {\bf B434} 54, (1998) (hep-th/9710112)
\bibitem{Zirnbauer} M.R.~Zirnbauer {\sl Conformal Filed Theory of the Integer Quantum
Hall Plateau Transition}, hep-th/9905054
\bibitem{WW} E.T.~Whittaker and G.N.~Watson, {\sl A Course of Modern Analysis}, C.U.P. (1940)
\bibitem{Shaharetal} D.~Shahar, D.~C.~Tsui, M.~Shayegan, E.~Shimshoni and
S.~L.~Sondhi, Phys. Rev. Lett. {\bf 79}, 479 (1997) (cond-mat/9611011)
\bibitem{Cage} M.M.~Cage in {\sl The Quantum Hall Effect}, Eds. R.E.~Prange and S.M.~Girvin, Springer, (1987)
\bibitem{Rendell+Girvin} R.W.~Rendell and S.M.~Girvin, Phys. Rev. {\bf B23}, 6610 (1981)
\bibitem{SHLTSR} D.~Shahar, M.~Hilke, C.C.~Li, D.C.~Tsui, S.L.~Sondhi and M.~Razeghi, Solid Sate Comm. {\bf 107}, 19 (1998) (cond-mat/9706045)
\end{thebibliography}
\end{document}